# A Novel Grid Based Dynamic Energy Efficient Routing Approach for Highly Dense Mobile Ad Hoc Networks

Baisakh[1], Chinmayee Mishra[2], Abhilipsa Pradhan[3]

Department of Computer Science and Engineering,
Sambalpur University Institute of Information Technology
`baisakh@suiit.ac.in`

**ABSTRACT**

*A research work without considering the power constraint cannot be conceded a fine contribution towards Mobile Ad Hoc Networks (MANET). As MANET comes into action for some special purpose, but its fugacity sometimes result degrades in network performance. Although the many prominent features of MANET like mobility, dynamic change in topology, multi radio relaying, quickly lay down the network without depending upon fixed infrastructures and many more provides tremendous flexibilities for the end user but challenges like limited power constraint, reliable data communication, band width utilization , network performance and throughput are still needed to be handle very sensibly. As limited battery power and inefficient routing protocol mechanism are high prone to network partition, in such case the network needs to be established more than once. Because communication establishment involves many costly operations like route discovery and route maintenance. The more the network partition the more the packet drops and packet loss which indeed requires a number of retransmission of packets, consuming network bandwidth as well as depleting battery power of individual nodes with a higher rate, which are the major destructive elements in network performance degradation as well as the major cause of reducing individual node's life time and network life time. So with all caveat in mind, we have proposed a novel **Grid Based Dynamic Energy Efficient Routing (GBDEER)** approach which is aimed to construct an energy efficient path from source to destination based on grid area, where each grid will have three deferent levels of transmission power. Every grid will have its own **grid supervisor node** who will take the responsibility during data communication, especially when the data is been passed through that specific grid. And keeping the dynamic nature of MANET in mind, we have also provide the feature of **grid subordinate node**, who will take the place of grid supervisor in case the supervisor is moving out of the grid area or running out of energy from certain threshold level. So we our proposed method not only establishes an energy efficient path but also concerned a dedicated path which can be used for data communication for a long period of time without any network partition. Hence this approach will be less prone to all those problems mentions above by the incorporating an efficient mobility handling mechanism.*

## 1. INTRODUCTION

A mobile ad-hoc network ( MANET) [1,2, 14,15] multi-hop wireless network where every node has a transceiver to communicate with its all neighbor node without the presence of any fixed infrastructure. In such scenario even if the destination node is not present directly within the transmission range of the source node but data packet can be sent through the intermediate nodes. So in MANET the whole routing mechanism has to be incorporate in every node. As such big responsibility like routing in the absence of base station lies on every individual node where the nodes are operated by limited battery power. And so it requires an effective use of limited battery power. Very often the improper use of limited battery power becomes the potential result for network partitions. Once a network partition occurs incentives and suitable actions taken by the individual node which indeed a costly operation. There may be some other reason which make also leads to network partition. One of the reasons is mobility, which is a special feature of MANET allowing nodes to move any direction at any point of time. And due to this feature dynamic topology change has become another serious issue in MANET.

So even though routing [3,4,5,6,17,18] has to be effective in the MANET but more than it an energy efficient routing is much more needed to avoid the problems of network partition, packet retransmission, bandwidth utilization, network lifetime and network throughput.

Mainly all the energy efficient routing protocols, MDSR [7],MEADSR[8],ECDSR[9], should be dealing with the typical power consumption which occurs due to various reasons. In MANET power consumption at each node is broadly divided into three phases. The first one is power consume during transmitting the data packet, second one is power consumed during receiving the data packets and the third one is when the node is idle, still it consume some amount of energy. Along with this power constraint in MANET there are some other issues which make difficulties while developing an energy efficient routing protocols. One of the reasons is mobility where no node can be fixed with its position through a complete data communication. More ever its hard for individual node to get global information about all other nodes present in the networks. It hinders to get updation from its neighbor nodes which are highly needed in a dynamic environment where the topology may be changed at any point of time.

In this paper, we have proposed a Novel Greed Based Dynamic Energy Efficient Routing protocol (GBDEER), where we have facilitated to the global information like position, distance, mobility, available battery power of individual node to all other nodes taking part in data communication. We assume that with this all global information the node will take the vice discussion according to the situation needed .GBDEER is mainly aimed to avoid to network partition as much as possible, so that the communication can be carried out for longer period of time. If in the worst situation if any network partition occurs then GBDEER HAS efficient survival approach to handle a situation.

The next section contains the overview routing protocols where various routing approaches have been briefly discussed.

## 2. IN OVERVIEWS OF ROUTING PROTOCOLS MANET

### 2.1 Overview

In the above section we have briefly discussed about the need and demand for an energy efficient routing protocols[13,14,19,20,21,24,25,26,27,33,35,36,39] in MANET. So we can derive the conclusion that he dynamic change of retransmission power may be an efficient solution in reducing power consumption at each node with respect to have fixed transmission power. On this basis of dynamic transmission power there are many routing protocols[12,22,23,34,37,38] have been proposed. We have also concerned about the node's life time as well as network life time. In the same time we are also concerned about the node which are suffering from low available residual energy and being an intermediate node in the data communication. Some time a node with low residual energy may lead to a serious network partition as there is more chances of draining out of energy from the node and the consequences may lead to node dead.

So, keeping all these issues in mind researchers have started working on various field of MANET to deal all this issues cautiously. Broadly these fields have been divided into two parts. One is topology based routing algorithm and other one is positioned based routing algorithm. Basically topology based routing algorithm use the information of various connection between each pair of nodes and analyze the link structure of every node in order to forward data packets. This approach further be classified into two different subgroup as proactive protocols and reactive protocols. In proactive protocols, a routing information is stored in the routing table and every node will have its own routing table. So any time a node moves from one location to another, the routing table needs to be updated. This approach leads to a huge overhead when the data traffic is slower than the mobility rate. So the reactive protocols [10] came into action where each node has a route cache. The route cache contains all the routing information about its all neighbor nodes. One of the efficient routing protocol which comes under reactive protocol is DSR [11]. But the major drawback of this protocol is that DSR works on the minimum hop count principle and it does not bother about the power consumption at every node. As the distance between every pair of node gets higher, the transmission power required for the communication will become more. And higher transmission power will always lead to greater power consumption.

In position based routing [28,30], data forwarding is based on geographical direction of the destination. GAF is the best example which comes under this category. As there is no flooding the power consumed by individual node becomes less and hence increase the lifetime of a node.

## 2.2 Geographical adaptive fidelity (GAF)

Geographical Adaptive Fidelity [29] is a location based routing protocol to estimate energy consumption which we can the location information needed. There is no addressing scheme like IP address. In GAF protocol each node uses location information based on GPS to associate itself with a " grid" which is based on the radio range (communication range of each node). The whole area is divided into number of square grids. The grid may or may not contain nodes. Within each grid the node which is having highest residual power [27] is known as the supervisor of the grid. Two or more nodes in a same grid must be equivalent when they maintain the same set of neighbor nodes and belongs to same communication route. Each node uses its GPS indicated location to associate itself with a point in the virtual grid.

The nodes belong to which area that area is considered as a network. Then we can divide the whole network as "virtual grid" means small square area as shown in figure:1. It based on the formula: $r <= R/\sqrt{5}$. The grid may or may not contain nodes. Within each grid the node which is having highest residual power is known as the supervisor node of the grid. Two or more nodes in a same grid must be equivalent when they maintain the same set of neighbor nodes and belongs to same communication route. Each node uses its GPS indicated location to associate itself with a point in the virtual grid. After finding the grid supervisor nodes we can find a lot of path between the sources to destination for communication. Now our work is to find the minimum path between sources to destination by using the parameter distance. To find these shortest paths we consider minimal spanning tree [31,32] After finding the shortest path the packet can be delivered from source to destination.

Consider a certain area, which having number of nodes randomly, each node have some communication range or radio range let it be R. Each node knows its own location as well as its direct neighbor location via GPS.

G A F  d e f i n e s  3  s t a t e s :-
- Discovery (for determining the neighbors in the grid)

- Active (reflecting participation in routing)

- Sleep (when the radio is turned off) (how depart time is calculated)

Before leaving from the grid each node estimates it's **depart time** of grid and send it to its neighbor. The sleeping neighbor adjusts their sleeping time accordingly in order to keeping routing fidelity.

Let consider the size of the grid as r. The grid size r can be easily deduced from the relationship between r and radio range of nodes, R which is given by the formula: r <= R/$\sqrt{5}$

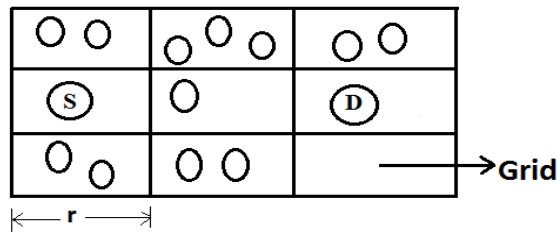

Figure:1Grid area creation

Our motive is to reduce power consumption in each node which extent the network life time. For this we proposed a new idea. In our work we first created virtual grid [1], which having a selected supervisor node along with number of common nodes. Taking consideration the supervisor node, we choose a appropriate shortest path from source to destination which based on minimum distance, to packet forwarding for better communication. We dynamically increase the transmission power. The transmission range can be increase or decrease according to the distance between nodes.

## 3. PROPOSED WORK

### 3.1 Overviews

In our proposed work, we are more concerned about a reliable communication where once a communication is established, should be carried out for a long period without any interruption. Here, reliable communication means the communication which does not involve too many route discoveries in a multipath system where more than one path is available between a pair of source and destination. We have strived to achieve it by choosing the best path from source to destination among all possible paths for the same pair of source to destination. As we all know that a good communication [16] in MANET should possess less power conservation, better throughput and low rate of packet retransmission. Because due to the dynamic change in topology, the dedicated path keeps changing with time and a special attention is required, so that the communication will get disturbed. Because a disturbed communication always requires additional energy to re-establish the path, retransmitting the packets for the same source to destination. As every participated node takes the responsibility of a complete successful communication, the limited battery power should be efficiently used, otherwise the node will die out of energy within a small span of time. These problems are given with many different solutions and still many research are going on to find best out of it. Well our work has merged two different ideas of GAF (Geographic Adaptive Fidelity) and minimum spanning tree.

- GAF (Geographic Adaptive Fidelity)
- minimum spanning tree

International Journal of Ad hoc, Sensor & Ubiquitous Computing (IJASUC) Vol.4, No.3, June 2013

By using the GAF we create many "virtual grids" which is based on the radio range of each node. All grids have been classified into different levels. These grids are nothing but the set of nodes present in the MANET. So according to this idea, all nodes of MANET be distributed to different grids according to their position. So when a source wants to communicate with a destination, then we will choose a supervisor node called as Grid Supervisor, from every grid. The selection of grid supervisor will be based on some criteria which has been explained in the next section. So every grid will contain a grid supervisor who takes the responsibility sending and receiving data during communication. If any circumstances a supervisor node is not ready to take provide its service as a supervisor node then, it will hand over its responsibility to its next deserving candidate called as subordinate node. Now we can derive a conclusion that every grid should have at least one supervisor node and one subordinate node. And all other nodes of a grid will remain in a sleeping state in order to save energy. So basically we are keening only one node as active at a time and all other are set to be in the sleeping mode.

Once the grids have been created and respected supervisor nodes are chosen then we have made used of spanning tree to find the shortest path from the source to destination. Once the intermediate node which is here supervisor nodes are selected, then we will set the transmission power dynamically for each supervisor node. As we know that in wireless communication major energy is consumed during transmitting a packet, then receiving a packet. Small amount of energy is also used when a node is not taking part in communication and is in a sleeping mode. Keeping this thing in mind, we tried to change the transmission power for each pair of nodes.

Even though we have used the concept of GAF protocol in our model, it's having a serious limitation. According to it, as the tree will be constructed from the grids, the highest or level of tree should not be more than 3.Because if the level will be more than 3 then during data transmission the energy dissipation will be more. We have augmented this GAF and constructed a tree more than level 3. If the level of tree will be more than 3 then the tree will be subdivided into two sub tree and the root of second sub-tree (a supervisor node for second grid type) will be connected with the one of the children of first sub-tree(A supervisor node of first grid type) as shown in figure:2 and 3.

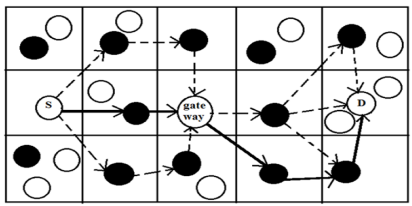

Figure:2 More than three level grid formation

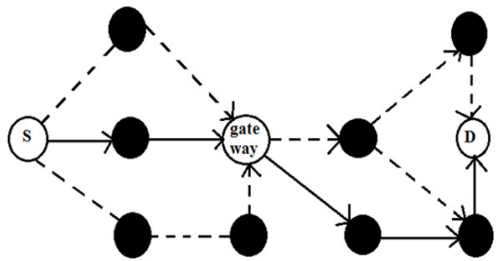

Figure:3 More than three level spanning tree

The detail has been explained in the next section. Our next section has also shaded some lights on the real technique behind the route discovery and route maintenance which are the two important phases of data communication in MANET.

- GBDEER ROUTE DISCOVERY
- GBDEER ROUTE MAINTAINANCE

### 3.2 GBDEER route discovery:

We have introduced tree levels of transmission power for different purposes and named it as $T_{max}$, $T_{mid}$ and $T_{min}$[35]. While calculating the grid area we have set the radio range R to Tamx. According to that we will get all different grids where grids at one level will be reachable to all girds at next level.The next step is to find the grid supervisor where the selection criteria will be maximum residual battery power and lowest node mobility. The node which is having highest weight based on this criteria will be the grid supervisor. Once the grid supervisor will be chosen then, the grid supervisor must know which node should be the next candidate for getting grid supervisor and that node is called as subordinate node. The subordinate node is also selected with the same criteria. Once the supervisor node and subordinate node are found then by using minimum spanning tree we will come into account to choose the best possible path from the same source and destination on the basis of shortest distance. After building the spanning tree, we have set a dynamic transmission power for each pair of grid supervisor nodes which are the part of minimum spanning tree shown in figure:4 and 5. It is possible by using the control packet named as Listen_ contrl. First the source node will broadcast the control packet to its neighbor grid supervisor which is present at the next level of grid. Initailly the Source node will set the transmission power to Tmin and with this transmission power a control packet will be broadcasted as shown in figure:6. If any node receives this control packet then, it will send a acknowledgement to the source node which is also a control packet with ACKListen_contrl with the same transmission power that is Tmin. If the source node will not get any ACKListen_contrl then it will increase its transmission power with the next transmission range that is Tmid.If after sending the control packet with tmid even, it won't get any ACKListen_contrl from its neighbor grid supervisor then, Tmax will be set. Once the transmission power is set been each pair of grid supervisors then data path communication will be established with required transmission power.

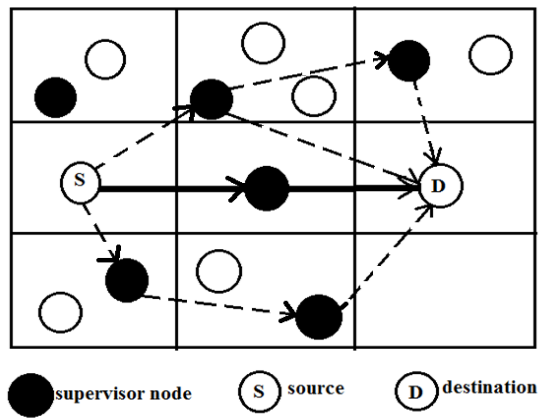

Figure:4 Grid creation with grid supervisor nodes

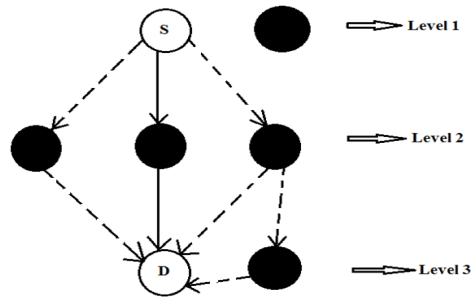

Figure:5 Spanning tree of grid supervisor nodes

### 3.2.1 Algorithm for GBDEER route discovery:

1. Initially set $T_{max}$ at each node and we find the complete network area where all any node can participate.

2. A "virtual grid area" is calculated and number of grids are formed.

3. A grid supervisor node is chosen on the basis highest residual energy and lowest mobility among all nodes present in a grid area.

4. Make a spanning tree of taking all grid supervisor node.

5. Set the transmission power at each supervisor node from all three different transmission levels like $T_{max}$, $T_{mid}$, $T_{min}$.

### 3.3 GBDEER Route Maintenance:

Another most important phase of network establishment is Route maintenance which is not only difficult operation to be performed but also a costly one which consumes energy with a greater rate. We have imposed an innovative idea that is the moment any grid supervisor is going out of its grid area or it's available energy is less than the specific threshold value then it's going to handover the responsibility to the subordinate node which belongs to the same grid area. The selection of the subordinate grid is same as the grid supervisor. In such scenario, the grid supervisor generates an error message to its entire neighbor node informing that he is no more a grid supervisor node rather the subordinate will become the grid supervisor node. In such way data can be continuously sent even though a new path is found for the same source to destination. And this special feature avoids the network partition by providing a new route.

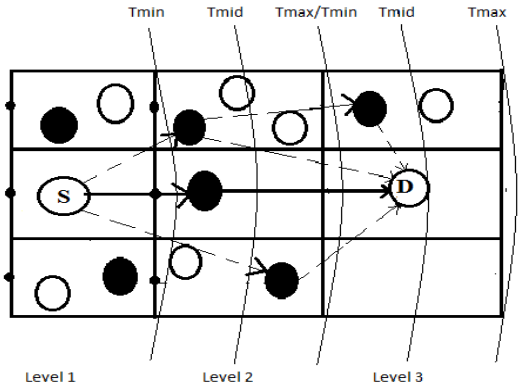

Figure:6 Three different transmission range $T_{max}$, $T_{mid}$, $T_{min}$

Another interesting feature has been added here that is dynamic change of transmission power[31][34] as the grid supervisor may be expected to move at any point of time. And any research work, certainly shouldn't avoid the mobility factor in MANET. So we have cover broadly two different possibilities that is a grid supervisor may move away from or move towards it's neighbor grid supervisor. If it comes closer towards its neighbor nodes, then according to the speed of the node the transmission power can be calculated which is dependent on the distance. Now the distance in such case can be calculated by $T_{trx} = T_{mid} - SPEED*TIME$. If we consider the second alternative case then as the distance between the pair of grid supervisor will be increased then the transmission power has to be increased by $T_{trx} = T_{mid} + SPEED*TIME$. In the worst case we may consider both the pair of nodes are moving away from each other and the distance will be

$$D = SPEED_s * TIME_s + SPEED_d * TIME_d$$

Where $SPEED_s$ is the speed of source node and $SPEED_d$ is the speed of the destination node with which it travels.

### 3.3.1 Algorithm for GBDEER route maintenance:

step -1: Initially assign a threshold value to each node.

Step-2: If value ((Supervisor Node $_{remaining\ residual\ energy}$) < (grid supervisor out of grid area)

then it must select subordinate nodes which is currently in a sleeping mode and set the transmission in which the respective grid supervisor in the begging.

Step-3: If the grid supervisor node move towards the source

then decrease the transmission by using the formula : $T_{trx} = T_{mid} - SPEED*TIME$

Step-4: If the node outwards to the range

then increase the transmission range to

$T_{trx} = T_{mid} + SPEED*TIME$

18

## 4   CONCLUSIONS

 In this paper we have propounded a new dynamic energy efficient routing protocol that is GBDEER which encompasses many special concepts like grid area, grid supervisor, grid subordinate, spanning tree and we have made an attempt to establish a reliable connection between a pair of source and destination by avoiding the chances of network partitions as much as possible. We believe that GBDEER will come out as a fair routing protocol by reducing the energy involved in additional route discovery, route maintenance. This may result less number of retransmission, packet loss, band width utilization, end to end delay and improve the network throughput and overall network performance. Although in practical, its quite challenging and difficult to get hold of all these performance based parameters together for providing a better result, but surely this approach can provide significantly good results in case of network partition, individual node's life time, network life time, reliable connection and many more. Mainly in GBDEER, the dynamic change in transmission power is highly responsible in prolonging network life time of the network, saving energy at every intermediate node and avoiding those node which are having less available battery power which tend to die soon. Our future work will be implementing GBDEER routing protocol and comparing with some of the other energy efficient routing protocols, taking all performance metrics need to be considered.

**Authors**

Short Biography

Mr. Baisakh is currently working as an Assitant Professor in Sambalpur Unioversity Institute of Information Technology since August,2012. He received his M.Tech degree in Computer Science from Jaypee University of Engineering and Technology, M.P., India in 2012 and B.Tech in Computer Science from Synergy Institute o Engineering and Technology, Odisha, India.

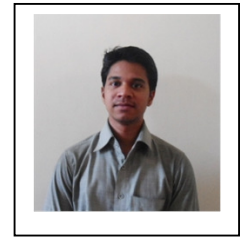